\documentclass[12pt]{iopart}
\usepackage{graphicx}
\usepackage{epstopdf}
\usepackage{color}
\usepackage{amsthm, amssymb, latexsym}
\usepackage{cite}
\usepackage{makecell}
\usepackage[section]{placeins}
\renewcommand{\figurename}{Figure}

\begin{document}

%\title[]{Parametric analysis of electron beam quality in a staged laser wakefield acceleration based on the ionization injection mechanism}

\title[]{Parametric analysis of electron beam quality in laser wakefield acceleration based on the truncated ionization injection mechanism}

\author{Srimanta Maity$^{*}$, Alamgir Mondal, Eugene Vishnyakov, Alexander Molodozhentsev$^{+}$}

\address{ELI Beamlines Facility, The Extreme Light Infrastructure ERIC, Za Radnicí 835, 25241 Dolní Břežany, Czech Republic \\}

\ead{$^{*}$srimantamaity96@gmail.com, $^+$alexander.molodozhentsev@eli-beams.eu}
\vspace{10pt}
%\begin{indented}
%\item[]October 2021
%\end{indented}
%~~~~~~~~~~~~~~~~~~~~~~~~~~~~~~~~~~~~~~~~~~~~~~~~~~~~~~~~~~~~~~~~~~~
\begin{abstract}

\noindent 
Laser wakefield acceleration (LWFA) in a gas cell target separating injection and acceleration section has been investigated to produce high-quality electron beams. A detailed study has been performed on controlling the quality of accelerated electron beams using a combination of truncated ionization and density downramp injection mechanisms. For this purpose, extensive two-dimensional Particle-In-Cell (PIC) simulations have been carried out considering a gas cell target consisting of a hydrogen and nitrogen mixture in the first part and pure hydrogen in the second part. Such a configuration can be realized experimentally using a specially designed capillary setup. Using the parameters already available in the existing experimental setups, we show the generation of an electron beam with a peak energy of 500-600 MeV, relative energy spread less than $5\%$, normalized beam emittance around 1.5 mm-mrad, and beam charge of 2-5 pC/$\mu$m. Our study reveals that the quality of the accelerated electron beam can be independently controlled and manipulated through the beam loading effect by tuning the parameters, e.g., laser focusing position, nitrogen concentration, and gas target profile. These simulation results will be useful for future experimental campaigns on LWFA, particularly at ELI Beamlines.

\end{abstract}

%~~~~~~~~~~~~~~~~~~~~~~~~~~~~~~~~~~~~~~~~~~~~~~~~~~~~~~~~~~~~~~~~~~~
%~~~~~~~~~~~~~~~~~~~~~~~~~~~~~~~~~~~~~~~~~~~~~~~~~~~~~~~~~~~~~~~~~~~
\section{Introduction}\label{intro}
Laser wakefield acceleration (LWFA)\cite{Tajima1979, RevModPhys.81.1229} is considered one of the most promising concepts to revolutionize particle accelerator technology. A typical laser-plasma accelerator (LPA) \cite{509991, bingham2003plasma, malka2005laser, joshi2007development, hooker2013developments} can offer acceleration gradients ($\sim$100~GV/m) more than three orders of magnitude larger than that achieved in a conventional radio-frequency (rf) accelerator ($\sim$100~MV/m). Thus, LPA can accelerate electrons to high energy (e.g., order of GeV) over a very short distance (a few centimeters), opening the path for developing compact electron accelerators.

The concept of particle acceleration using plasma as an accelerating medium was initially introduced in 1956 in three independent research works carried out by Budker \cite{Budker:1241561}, Veksler \cite{Veksler:1241563}, and Fainberg \cite{Fainberg:1241564}. Later in 1979, in a pioneering theoretical work \cite{Tajima1979}, Tajima and Dawson proposed how an intense laser pulse can excite large-amplitude plasma wakes and eventually accelerate electrons to high energy. The first experimental demonstration of electron acceleration using a laser-plasma system was reported in 1993 in Ref. \cite{PhysRevLett.70.37}. Since then, significant progress has been made to experimentally demonstrate a high acceleration gradient generating high-quality monoenergetic electron beam using laser-plasma interactions \cite{everett1994trapped, modena1995electron, mangles2004monoenergetic, faure2004laser, geddes2004high, wiggins2010high, pollock2011demonstration}. A breakthrough was achieved in 2006 \cite{leemans2006gev} experimentally demonstrating GeV electron beams using a centimeter-scale pre-formed capillary plasma channel. The latest achievement was reported in 2019 \cite{PhysRevLett.122.084801} demonstrating a quasi-monoenergetic electron beam with 7.8 GeV peak energy in a capillary discharge
waveguide using a Petawatt Laser. However, the generation of high-quality electron beams, compared with the conventional rf accelerator, remains challenging for a plasma-based accelerator. 

A significant effort is being made to improve the quality, e.g., high charge, low energy spread, small beam emittance, and low divergence of the accelerated electron beam. The electron beam quality in a laser-plasma accelerator (LPA) strongly depends on the injection mechanism. Over the past few decades, many injection schemes, e.g., self-injection \cite{bulanov1992nonlinear, xu2005electron, ohkubo2006wave, oguchi2008multiple, PhysRevLett.103.135004, PhysRevLett.103.175003, PhysRevLett.103.215006, PhysRevSTAB.15.011302}, density down-ramp injection \cite{PhysRevE.58.R5257, PhysRevLett.86.1011, PhysRevLett.100.215004, PhysRevSTAB.13.091301, hue2023control, fourmaux2012quasi, PhysRevSTAB.16.011301, PhysRevLett.110.185006}, colliding pulse injection \cite{PhysRevLett.79.2682, PhysRevE.59.6037, Kotaki10.1063, faure2006controlled, Kotaki4599030, PhysRevLett.102.164801, PhysRevLett.103.194803}, and ionization injection \cite{10.1063/1.2179194, PhysRevLett.98.084801, PhysRevLett.104.025003, PhysRevLett.104.025004, PhysRevLett.105.105003}, have been proposed, investigated, and demonstrated in simulation and experimental studies. Out of these methods, the ionization-induced injection mechanism is considered one of the most promising techniques for controlled injector optimization. This is because it is an easy-to-implement technique to realize in experiments, and a large set of parameters can be used to externally control the electron injection and the quality of the accelerated beam. The ionization injection scheme utilizes the large ionization threshold of the $K$-shell electrons of a high-$Z$ gas (e.g., nitrogen, oxygen) mixed with a low-$Z$ gas (e.g., hydrogen, helium) to control the initial phase of the injected and trapped electrons. The leading edge of the laser pulse is intense enough to fully ionize the low-$Z$ gas atoms along with the outer-shells electrons of the high-$Z$ gas atoms. The laser ponderomotive force pushes these pre-ionized electrons away and excites wakes. However, the $K$-shell electrons of high-$Z$ gas atoms are only ionized near the peak of the laser profile and, thus, slip backward relative to the phase velocity of the wake. Under certain conditions \cite{PhysRevLett.104.025003}, these electrons are trapped and accelerated further as they move forward with respect to the wake. Thus, using high-$Z$ gas atoms provides external control over the electron injection, allowing electron trapping even at lower plasma densities and with lower laser intensities compared to the self-injection scheme. However, the energy spread of the accelerated electron beam produced using this scheme is typically very large because of the continuous injection of electrons until the end of the mixed gas or until some physical mechanism, e.g., beam loading, pump depletion, stops injection \cite{PhysRevAccelBeams.21.052802}. On the other hand, the density downramp injection scheme relies on the decrease in the wake phase velocity and wake wave breaking that occurs due to the inhomogeneity of the Langmuir frequency \cite{PhysRevE.58.R5257}. Thus, this technique can be used for localized injection of electrons in the wake structure, reducing the energy spread of the accelerated electron beam. In this study, we have demonstrated that ionization and density downramp injection mechanisms can be compatible with each other, and their combined effect can produce high-quality, high-energetic electron beams.

Recent works \cite{irman2018improved, PhysRevLett.126.104801, PhysRevLett.126.174801, PhysRevLett.129.094801} have demonstrated that ionization-induced injection length can also be controlled, producing a quasi-monoenergetic electric beam. For example, using an initially unmatched laser pulse (i.e., $k_pw_0 \neq 2\sqrt{a_0}$, where $k_p$, $w_0$, and $a_0$ represent wave number of a plasma wave, laser spot size, and amplitude of the normalized vector potential of the laser pulse) automatically truncates the injection process due to the violation of the ionization injection conditions. This technique has been termed as 'self-truncated ionization injection (STII)' \cite{zeng2014self, mirzaie2015demonstration, li2020control}, which utilizes the relativistic self-focusing effect \cite{10.1063/1.866349, PhysRevLett.70.2082, PhysRevLett.74.710} of the laser pulse. In several simulations \cite{PhysRevLett.126.104801, PhysRevLett.126.174801,PhysRevLett.129.094801} and experimental studies \cite{PhysRevLett.107.035001, PhysRevLett.107.045001, PhysRevLett.111.165002}, the use of a gas target consisting of a few mm-long mixed gas (doped with high-$Z$ gas) region followed by a large volume of a pure gas (low-$Z$) has also been proposed and demonstrated to reduce the injection length without compromising the acceleration length. The influence of a chirped laser pulse on the electron trapping and the quality of the electron beam in an LWFA operating in the bubble regime has been investigated in the past \cite{deutsch1991strong, kalmykov2012laser, ghotra2022laser}. An analytical study suggesting the scaling and design parameters for generating a multi-pico Coulomb and multi-GeV electron beam from LWFA with a cm-scale gas cell has also been reported recently \cite{ghotra2023multi}. However, a systematic study showing a direct dependence of various system parameters on the accelerated electron beam quality remains unexplored.   

Our current work investigates the capabilities to produce high-quality electron beams in a gas cell target in the framework of laser wakefield acceleration. For this purpose, a systematic and comprehensive study using two-dimensional PIC simulations has been carried out. The gas cell target considered in our study consists of two sections. In the first section, a truncated ionization-induced injection occurs where a short region of nitrogen-doped hydrogen is considered. In the second section, the injected electrons are accelerated in a long plasma plateau formed from pure hydrogen. The dependence of laser parameters and density profile on the quality of accelerated electron beams has been explored. In particular, we have explicitly shown the effect of laser focusing position, nitrogen concentration, and the initial density profile of the gas target on the injection process as well as on the properties of the accelerated electron beam. Our study affirms how these parameters can be used to independently control and manipulate several beam properties, e.g., beam charge, peak energy, emittance, and beam divergence. 

The remaining part of this paper has been organized as follows. In section \ref{simu}, PIC simulation details and simulation setup highlighting the parameters used in this study have been discussed. A detailed discussion of the observations obtained from simulations is provided in section \ref{rd}. The subsections therein contain the simulation results for various system parameters, e.g., laser focusing position, concentration of nitrogen, and target density profile. A summary of our study, including the conclusive remarks, has been provided in section \ref{summary}.

%~~~~~~~~~~~~~~~~~~~~~~~~~~~~~~~~~~~~~~~~~~~~~~~~~~~~~~~~~~~~~~~~~~~
%~~~~~~~~~~~~~~~~~~~~~~~~~~~~~~~~~~~~~~~~~~~~~~~~~~~~~~~~~~~~~~~~~~~   
\section{Simulation Setup}\label{simu}

A set of two-dimensional (2D) Particle-In-Cell (PIC) simulations has been performed to investigate the dependence of electron beam properties on various system parameters in a laser wakefield accelerator (LWFA). For this purpose, a fully relativistic, massively parallel, open source PIC code, EPOCH 4.18.0 \cite{arber2015contemporary, bennett2017users} has been used. A 2D simulation box with size $80$-$\mu$m $\times$ $160$-$\mu$m in the x-y plane is considered. The cell size is considered to be $0.02$-$\mu$m $\times$ $0.08$-$\mu$m, which corresponds to $4000$ and $2000$ number of grid cells in the $x$ and $y$ directions, respectively. We used eight simulation particles in each cell initially. In EPOCH, the standard Boris algorithm along with a modified leapfrog method has been used for particle pusher. The details about this code can be found at the Ref. \cite{arber2015contemporary}. The moving window scheme has been utilized in our study so that a relatively very small simulation box can be used to simulate a large acceleration distance. The open boundary condition has been used for both electromagnetic (EM) waves and particles, where EM waves and particles are transmitted through the boundary and removed from the system. The time interval between the two steps is chosen following the CFL (Courant-Friedrichs-Lewy) condition ($C =  c\Delta t /\Delta x + c\Delta t /\Delta y \leq 1$). Here $c$ is the speed of light in the vacuum, and the dimensionless number $C$ is called the Courant number. In our study, we have chosen the value of $C$ to be $C=0.99$ and, thus, simulation time step $\Delta t =0.052$ fs. To resolve all the physical processes associated with the wakefield acceleration, the Courant number $C$, following the CFL condition, should be as close as 1.0, and the grid size ($\Delta x$) along the laser propagation direction should be less than $\lambda_L/30$, where $\lambda_L$ represents laser wavelength. In our simulations, we have taken care of these conditions.
A Gaussian laser pulse with wavelength $\lambda_L = 800$ nm and a Full-Width-Half-Maximum (FWHM) pulse duration $\tau_{fwhm} = 30$ fs is used in this study. The laser also has a Gaussian transverse profile (along $y$) and is focused at a $f$ distance (which will be varied in our study) along $\hat x$ from the left boundary of the simulation box. The Full-Width-Half-Maximum (FWHM) of laser spot size $w_{fwhm}$ at focus is considered to be 30 $\mu$m, which corresponds to the width (radius) of the laser spot at focus, $w_0 \simeq 15$ $\mu$m. The initial laser energy (at focus) is considered to be $1.5$ J, which corresponds to the peak power ($P_{0}$) of the laser of $50$ TW and a normalized vector potential $a_0$ around 2.0 at focus. Table \ref{table1} also provides all the relevant simulation parameters.

In our simulation study, we have used a combination of nitrogen (N$_2$) and hydrogen (H$_2$) gas as a target. The schematic of the simulation configuration and on-axis density profile of the gas mixture has been depicted in Fig. \ref{dens_ini}. A detailed discussion related to the gas target configuration is provided in Sec. \ref{rd}.

\begin{table}
\caption{Simulation parameters used in this study.}
\vspace{0.5cm}
\label{table1}
\resizebox{\textwidth}{!}{%
\begin{tabular}{|c||c||c||c|}
\hline
    \makecell{Wavelength\\ ($\lambda_L$)} & \makecell{Spot size (FWHM)\\ ($w_{fwhm}$)} & \makecell{Pulse duration (FWHM)\\ ($\tau_{fwhm}$)} & \makecell{Laser peak power ($P_0$)}\\ 
\hline
    0.8 $\mu$m & 30 $\mu$m & 30 fs & $50$ TW\\ 
\hline
\hline
\hline
    \makecell{Laser frequency\\ ($\omega_L$)} & \makecell{Laser energy\\ ($E_L$)} & \makecell{Peak value of normalized\\ vector potential ($a_0$)} & Peak intensity $I_0$ (W/cm$^2$)\\ 
\hline
    $2.35\times 10^{15}$ Hz & 1.5 J & 2.1 & $1.0\times 10^{19}$\\     
\hline
\hline
\hline
\makecell{Hydrogen density in the plateau \\ ($n_{Hp}$)} & \makecell{Plasma frequency\\ ($\omega_{pe}$) corresponding to $n_{Hp}$} & \makecell{N$_2$ concentration\\ ($C_N$)} & $\omega_L/\omega_{pe}$\\ 
\hline
    $1.6\times 10^{18}$ cm$^{-3}$ & $7.13\times 10^{13}$ Hz & $1-17\%$ & 33\\
\hline        
\end{tabular}}
\end{table}

%~~~~~~~~~~~~~~~~~~~~~~~~~~~~~~~~~~~~~~~~~~~~~~~~~~~~~~~~~~~~~~~~~~~
%~~~~~~~~~~~~~~~~~~~~~~~~~~~~~~~~~~~~~~~~~~~~~~~~~~~~~~~~~~~~~~~~~~~
\section{Simulation Results}
\label{rd}

\begin{figure}
  \centering
  \includegraphics[width=5.2in,height=2.5in]{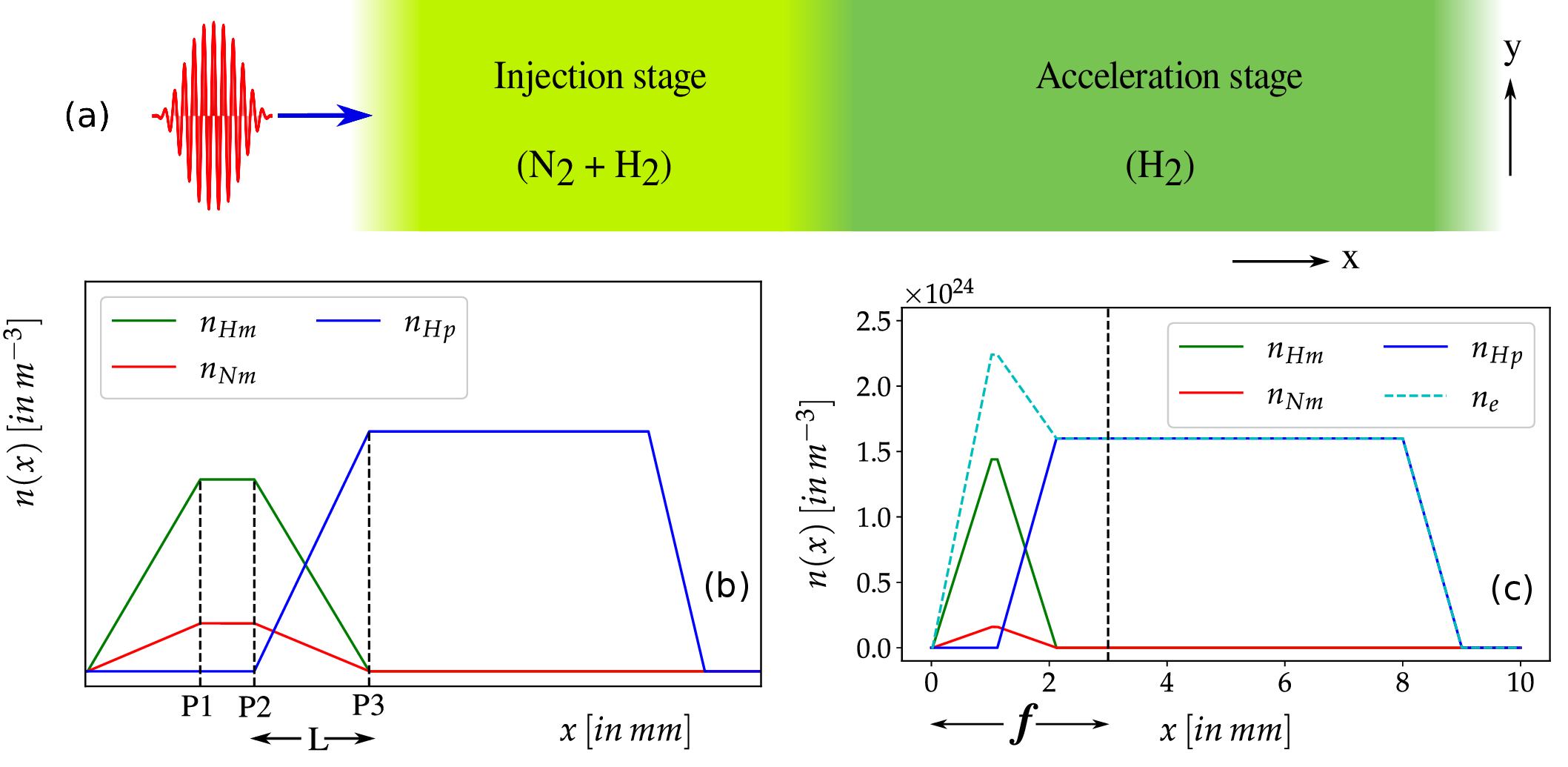}
  \caption{(a) A schematic of the `dual-stage' configuration of LWFA considered in our simulation study has been shown. In the first section of the target (injection stage), a mixture of nitrogen (N$_2$) and hydrogen (H$_2$) gas has been considered. In the second section (acceleration stage), only pure hydrogen (H$_2$) is considered. (b) A schematic of the 1D density profiles of H$_2$ and N$_2$ along $x$ is shown. Here, $n_{Hm}$, $n_{Nm}$ represent the hydrogen and nitrogen neutral (atomic) density in the first section, respectively and $n_{Hp}$ represents the (atomic) density of the pure hydrogen in the second section of the gas cell target. (c) Variations of $n_{Hm}$, $n_{Nm}$, $n_{Hp}$ and corresponding net electron density $n_{e}$ along $x$ at $y = 0$ are shown for a particular case. Here, in this particular case, ($90\% + 10\%$) N$_2$-H$_2$ mixture has been used and the values of P1, P2, and P3 are chosen to be $1.02$ mm, $1.12$ mm, and $2.12$ mm, respectively. The black dashed line at a distance $f$ from the left boundary of the simulation box represents the laser focusing position.}
\label{dens_ini}
\end{figure}

In this study, the ionization-based injection mechanism and its dependence on various system parameters of an LWFA have been explored in detail. In particular, a dual-stage configuration of LWFA has been considered, as illustrated by the schematic in subplot (a) of Fig. \ref{dens_ini}. In the first section of this configuration, a gas mixture of nitrogen (N$_2$) and hydrogen (H$_2$) has been considered. This section ranges over a shorter distance ($\sim 2$ mm) and has been termed as \textit{Injection stage}. As the laser propagates through this stage, the leading edge of the laser pulse ionizes the neutral hydrogen atoms and the outer five electrons of neutral nitrogen atoms. Thus, these electrons immediately start to experience laser ponderomotive force and are pushed out in both longitudinal and transverse directions to excite plasma wake. However, two K-shell electrons of each nitrogen atom having higher ionization potentials are ionized near the peak of the laser pulse profile and injected into the electric field of the fully formed plasma wake. The second section of the target, termed as \textit{Acceleration stage}, covers a larger distance ($\sim 6$ mm) and comprises pure hydrogen (H$_2$) gas. In this section, no further injection of electrons occurs. This section only provides the accelerating electric field as the laser pulse propagates through it to further accelerate the trapped electrons \cite{PhysRevLett.96.014803} entering from the \textit{Injection stage}. The gas target configuration considered in our study can be realized in experiments using a specially designed single capillary setup with an axial length of around 10-15 mm and a radius of 300-500 $\mu$m. A similar concept was recently implemented \cite{PhysRevLett.126.104801, PhysRevLett.126.174801, PhysRevLett.129.094801} to experimentally demonstrate the generation of high-quality electron beams.

A schematic of the 1D density profile along $x$ considered in this study has been demonstrated in Fig. \ref{dens_ini}(b). Here, $n_{Hm}$ and $n_{Nm}$ represent the atomic density of H$_2$ and N$_2$ in the first stage (\textit{Injection stage}), respectively. The parameter $n_{Hp}$ defines the atomic density of H$_2$ in the second stage, i.e., \textit{Acceleration stage}. In our study, we have also investigated the dependence on the initial density profile in the first stage by changing the values of $L$ (distance between the positions P3 and P2). The 1D density profile along $x$ for $L = 1.0$ mm has been shown in Fig. \ref{dens_ini}(c), where we have considered the atomic density of H$_2$ (hence the electron density) in the plateau region of the second stage to be $1.6 \times 10^{18}$ $cm^{-3}$. In our simulations, we have varied the three system parameters, i.e., laser focusing position ($f$), $N_2$ concentration ($C_N$), and initial density profile (i.e., changing values of $L$). We have analyzed various features of our observation, focusing on the dependence of electron beam properties on these system parameters, and presented them in the following subsections.

%~~~~~~~~~~~~~~~~~~~~~~~~~~~~~~~~~~~~~~~~~~~~~~~~~~~~~~~~~~~~~~~~~~~

\subsection{Effect of Laser Focusing Position}\label{effect_fc}

\begin{figure}
  \centering
  \includegraphics[width=5.8in,height=4.5in]{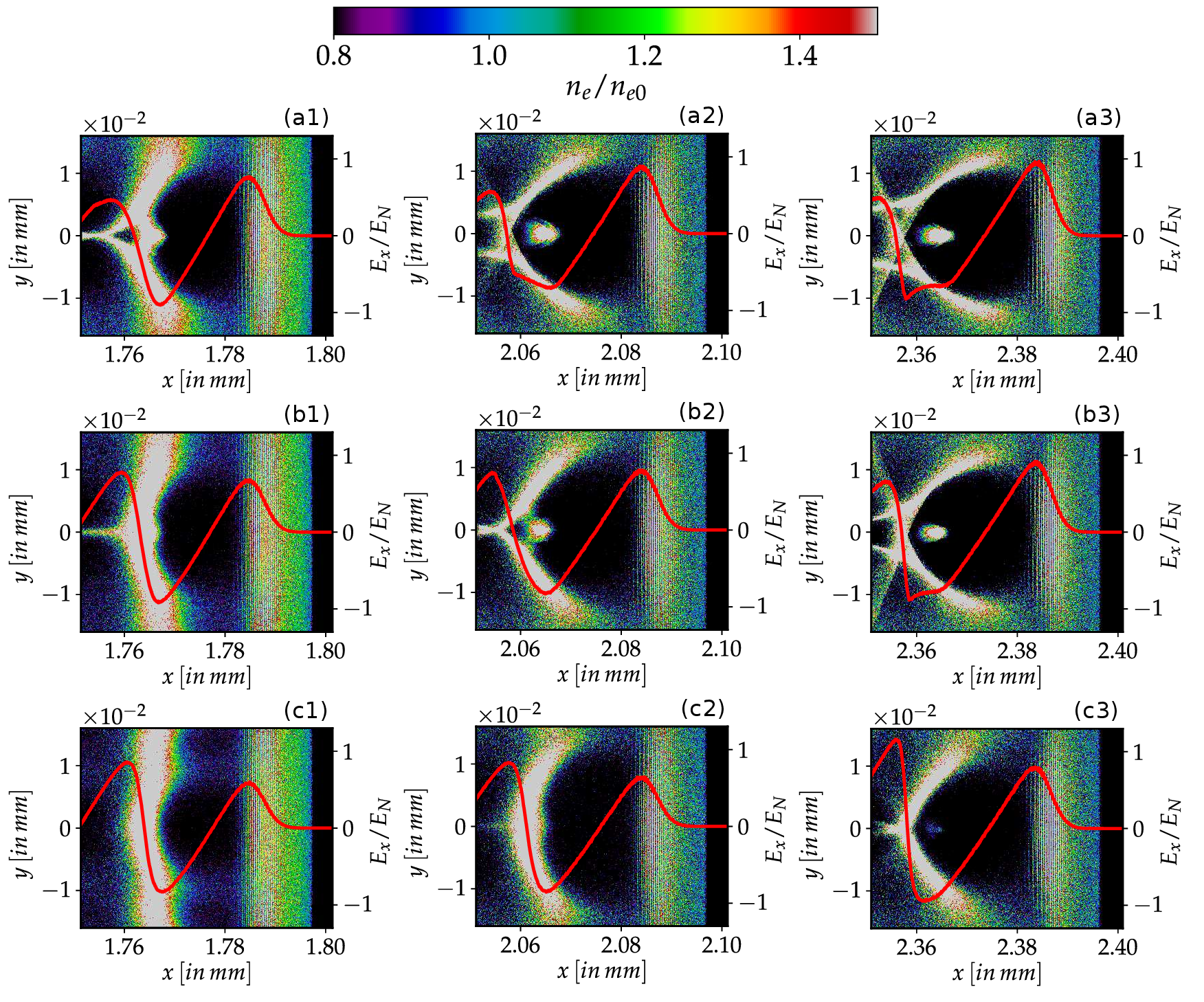}
  \caption{Distributions of electron density in the x-y plane at three different instants of time with changing values of laser focusing position ($f$) have been illustrated. In subplots (a1) $\omega_{pe}t = 430$, (a2) $\omega_{pe}t = 500$, and (a3) $\omega_{pe}t = 570$ laser focusing position is considered to be $f = 3.0$ mm. Whereas, in subplots (b1)-(b3) and (c1)-(c3), the laser focusing positions are considered to be $f = 3.3$ mm and $3.7$ mm for the same instants of time, i.e., $\omega_{pe}t = 430$, $500$, and $570$, respectively. Here, red solid lines in each subplots represent the on-axis longitudinal electric field ($E_x$) for the corresponding cases. Here, $n_{e0}$ is the electron density in the plateau regime of the second stage. In these simulations, N$_2$ concentration is chosen to be $C_N = 10\%$.}
\label{2d_dens_fc}
\end{figure}

\begin{figure}
  \centering
  \includegraphics[width=5.3in,height=4.0in]{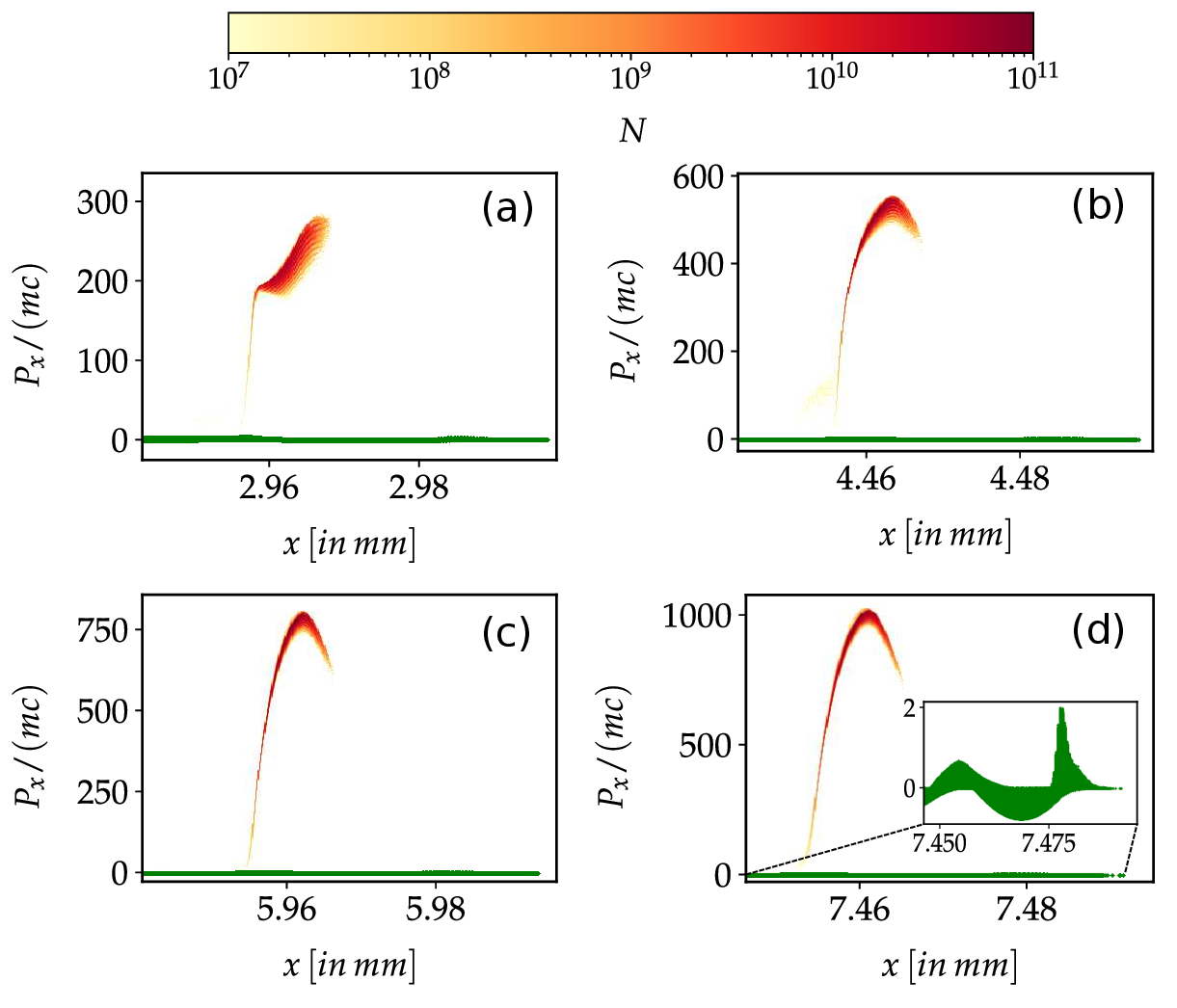}
  \caption{Longitudinal phase-space ($P_x$-$x$) distributions of electrons for a particular simulation run with $f = 3.5$ mm and $C_N = 10\%$ at time (a) $\omega_{pe}t = 715$, (b) $\omega_{pe}t = 1070$, (c) $\omega_{pe}t = 1425$, and (d) $\omega_{pe}t = 1780$ have been shown. The electrons originated from H$_2$ are represented by green markers. Whereas, electrons ($N$) from N$_2$ are depicted by pseudo-color symbols from yellow to red as shown by the color-bar.}
\label{px_fc}
\end{figure}

\begin{figure}
  \centering
  \includegraphics[width=5.2in,height=4.0in]{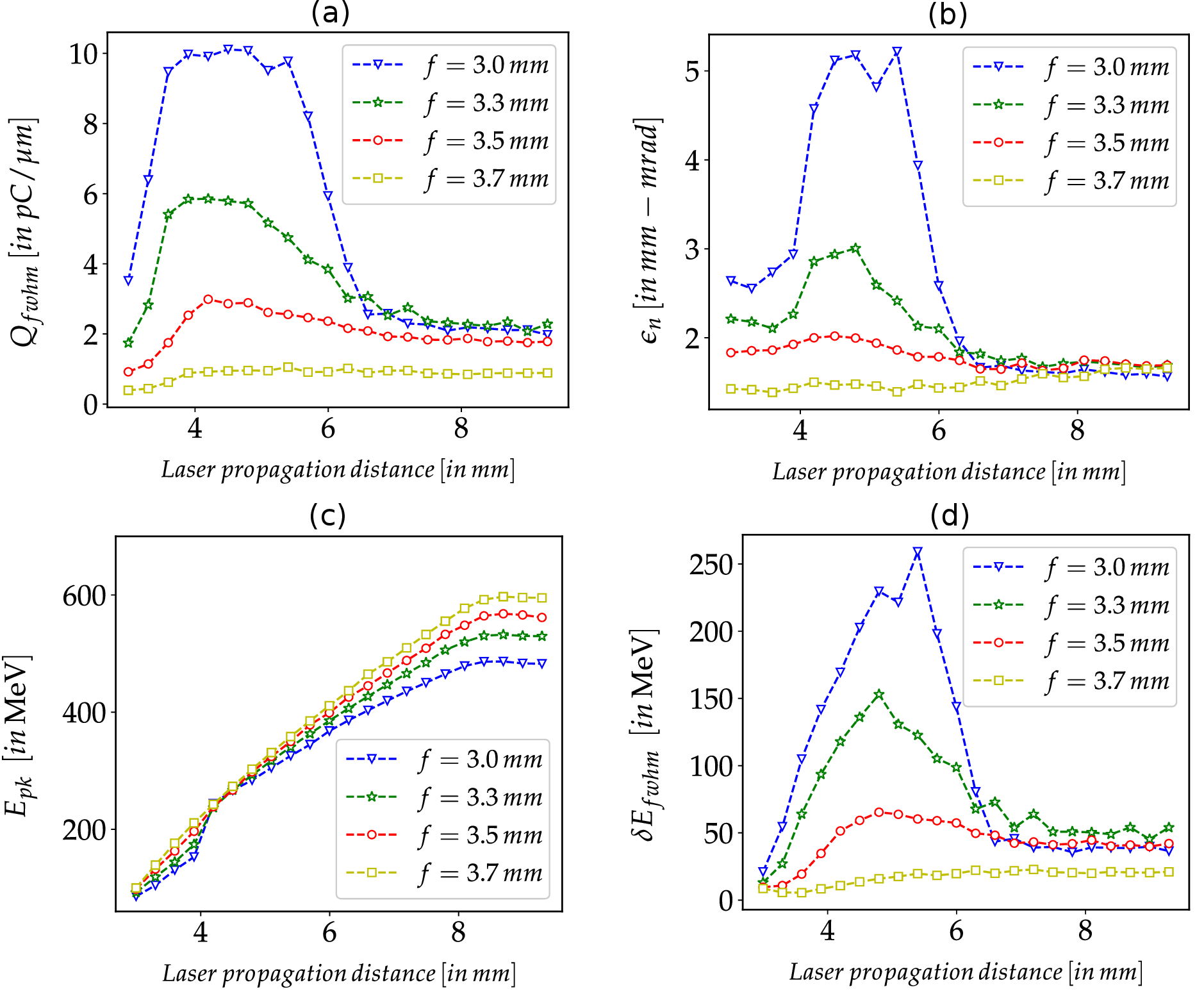}
  \caption{ Properties of electrons injected and trapped in the plasma wake structures behind the laser pulse as a function of laser propagation distance is shown for different values of laser focusing position ($f$). In subplots (a) and (b), the absolute values of charge ($Q_{fwhm}$) located within FWHM of energy histogram of the trapped electron beam and corresponding normalized beam emittance ($\epsilon_n$) are shown, respectively. Energy peak values ($E_{pk}$) and corresponding energy spread ($\delta E_{fwhm}$) of the beam are depicted in subplots (c) and (d), respectively.}
\label{beam_time_fc}
\end{figure}

\begin{figure}
  \centering
  \includegraphics[width=5.5in,height=4.0in]{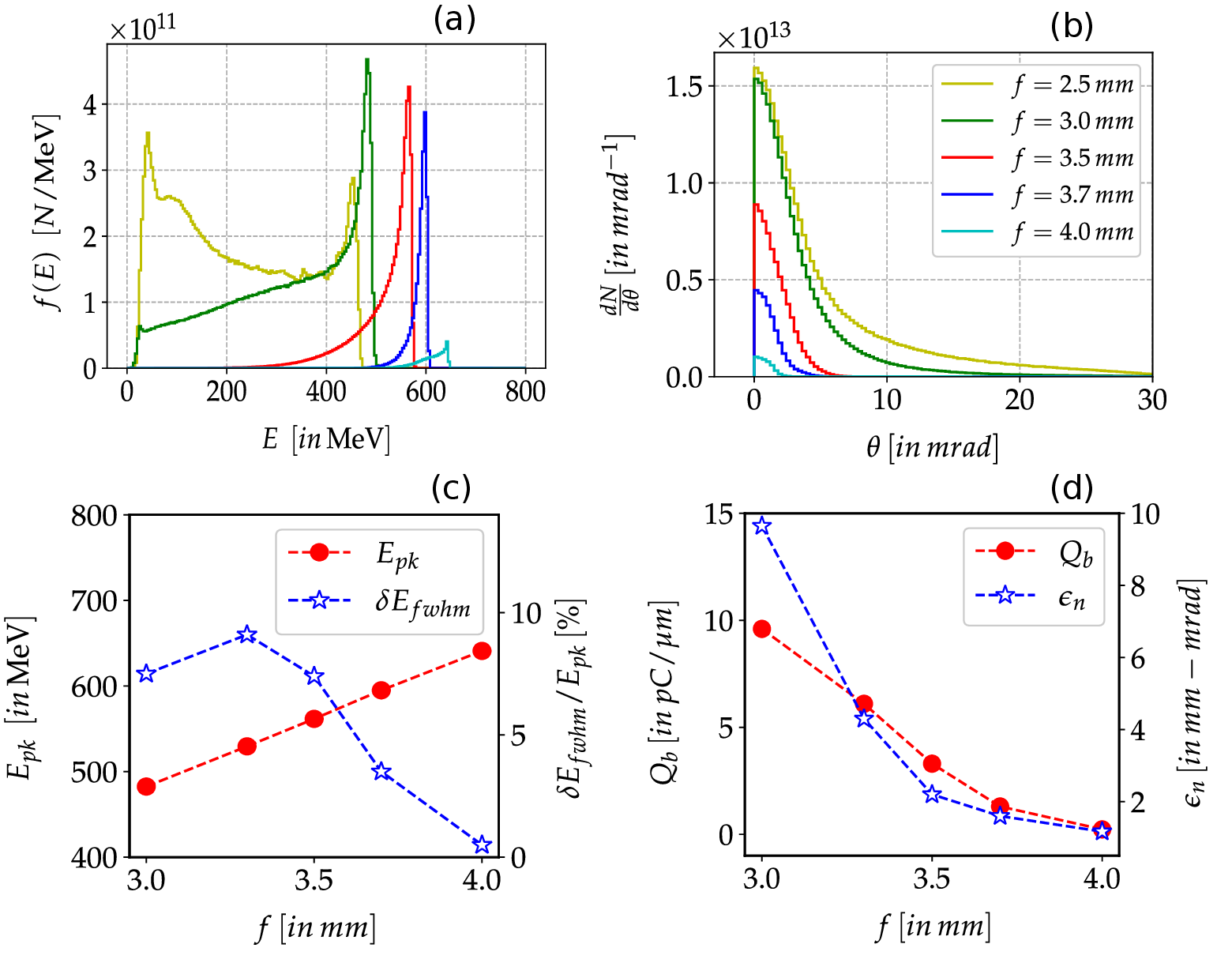}
  \caption{The histograms of electron beam energy and beam divergence at a particular time $\omega_{pe}t = 2212$ (i.e., laser after exiting from the plasma) for different simulations with changing values of $f$ are shown in subplots (a) and (b) respectively. The peak value energy ($E_{pk}$), energy spread ($\delta E_{fwhm}$) of the beam, net charge ($Q_b$), and normalized beam emittance $\epsilon_n$ (considering all the trapped electrons) at the same instant of time with changing values of $f$ are shown in subplots (c) and (d), respectively. }
\label{beam_fc}
\end{figure}

To study the effect of laser focusing position on the properties of the accelerated electron beam, a set of PIC simulations with changing values of $f$ has been performed. The laser focusing position has been varied from $f = 2.5$ mm to $4.0$ mm. In these simulations, ($10\% + 90\%$) N$_2$-H$_2$ mixture has been used in the first stage (\textit{Injection stage}) with a plateau density of atomic hydrogen $n_{Hp} = 1.6 \times 10^{18}$ $cm^{-3}$ in the second stage (\textit{Acceleration stage}). The first stage extends up to $x = 2.12$ mm, from where the plateau regime of the second stage starts. In our study, we have also simulated the cases with $f<2.5$ mm and $f>4.0$ mm. However, we have observed that for $f<2.5$ mm, although the net charge of the injected electron bunch becomes very high, the quality of the accelerated beam is very poor. Whereas, for $f>4.0$ mm, no significant injection of electrons was observed for our chosen values of system parameters. 

 Net electron density ($\rho$) distributions in the $x-y$ plane have been depicted in Fig. \ref{2d_dens_fc} with changing values of $f$ at three particular time instants. In subplots (a1)-(a3) of Fig. \ref{2d_dens_fc}, we have considered three different instants of time, i.e., $\omega_{pe}t = 430$, $500$, and $570$ with laser focusing position fixed at $f = 3.0$ mm, respectively. Whereas, for subplots (b1)-(b3) and (c1)-(c3) of Fig. \ref{2d_dens_fc}, laser focusing positions are considered to be $f = 3.3$ mm and $3.7$ mm for the same instants of time, respectively. It is seen that, in all cases, electrons get injected and trapped into the plasma wake. The injection process starts approximately at $x\sim 1.76$ mm, as can be seen from subplots (a1)-(c1) of Fig. \ref{2d_dens_fc}. It is clear that in all the cases, injections and trapping of electrons occur only within the density down-ramp regime of the first stage. Thus, in our case, both ionization injection and density down-ramp mechanisms effectively contribute to the injection and trapping of electrons. However, the timings of injection and trapping of electrons get shifted with the increase of $f$, as seen from subplots (a2)-(c2) Fig. \ref{2d_dens_fc}. The on-axis longitudinal electric field ($E_x$) profile shown by red solid lines for different values of $f$ at different times also manifests these phenomena. Fig. \ref{2d_dens_fc}(a2) shows that the profile of $E_x$ gets modified (around $x = 2.065$ mm) for $f = 3.0$ mm at this instant of time. This clearly indicates the beam loading effect \cite{wilks1987beam}, which occurs when electrons get trapped within the first quarter of $E_x$. However, at the same time, for $f = 3.3$ mm and $3.7$ mm, no effect of beam loading is observed, as can be seen from subplots (b2) and (c2) Fig. \ref{2d_dens_fc}, respectively. Finally, at time $\omega_{pd} = 570$ (Fig. \ref{2d_dens_fc}(b3)), the effect of beam loading starts to reveal for the case with $f = 3.3$. It is to be noticed that by this time, the laser pulse, as well as trapped electrons, had already entered the plateau regime ($x>2.12$ mm) of the second stage. However, beam loading does not occur for the case with $f = 3.7$ mm, as shown in Fig. \ref{2d_dens_fc}(c1)-(c3). The effect of beam loading on the accelerated electron bunch's quality will be discussed in section \ref{effect_cn}.

 The longitudinal phase-space ($P_x$-$x$) distributions of electrons at different times of the simulation run with $f = 3.5$ mm have been depicted in subplots (a)-(d) of Fig. \ref{px_fc}. Here, green markers represent the electrons originating from hydrogen, whereas yellow to dark red color symbols are for electrons ionized from nitrogen neutrals. As mentioned before, no nitrogen gas is present in the second stage ($x>2.12$ mm). Thus, it is clear that only the electrons ionized from the nitrogen atoms in the first stage get trapped and are accelerated further. It also reveals that no self-injection of electrons originating from H$_2$ occurs during the propagation of the laser through the second stage. Electrons from H$_2$ only form wake structures and provide the accelerating field, as shown in the inset of Fig. \ref{px_fc}(d).

We have characterized several properties of these trapped electrons as they travel through the second stage for different simulation setups with changing values of $f$. We have presented this in subplots (a)-(d) of Fig. \ref{beam_time_fc}. In our simulations, we have observed that the net charge of the trapped electron beam almost remains constant throughout the propagation. This indicates that the injected electrons do not get lost during their propagation through the second stage after being trapped in the wake structure in the first stage. However, the net charge located within the full width at half maximum (FWHM) of the energy histogram of the trapped electron beam changes with the laser propagation distance. In Fig. \ref{beam_time_fc}(a), we have depicted the absolute value of FWHM charge ($Q_{fwhm}$) of the trapped electrons for different values of $f$. For $f\leq 3.5$ mm, $Q_{fwhm}$ first increases with the propagation distance. However, after a certain distance, the value of $Q_{fwhm}$ decreases again and finally becomes constant at a certain value. For $f\geq 3.7$ mm, $Q_{fwhm}$ increases first as the laser propagates and finally gets saturated after a certain distance. 

The normalized beam emittance ($\epsilon_n$) as a function of time presented in Fig. \ref{beam_time_fc}(b) shows a strong dependence on the laser focusing position. Here, we have calculated $\epsilon_n$ using the transverse phase-space coordinates ($P_y$-$y$) of the particles located within the FWHM of the energy histogram. For lower values of $f$ ($f<3.5$ mm), beam emittance initially increases with the propagation distance, attains a maximum value, and decreases again as the beam propagates further. However, for all the cases, the peak energy ($E_{pk}$) of the trapped electron beam increases almost linearly with the laser propagation distance. Finally, it gets saturated at certain values, as seen in Fig. \ref{beam_time_fc}(c). This is consistent with the previously reported theoretical scaling \cite{lu2007generating} where the energy gain of an electron beam in an LWFA is linearly proportional to the accelerating distance within the dephasing length. The energy gain gets saturated after a certain distance for all the cases. This corresponds to the time when the laser pulse and trapped electron beams exit the plateau region of the second stage. It is also seen that the final saturated peak energy of the beam is higher for the lower value of the laser focusing position. In Fig. \ref{beam_time_fc}(d), the energy spread ($\delta E_{fwhm}$) measured from the FWHM of energy histogram of the trapped electron beam is shown for different simulation runs with changing values of $f$. We have observed that the energy spread does not remain constant, at least up to a certain time of evolution, and has a strong dependence on the laser focusing position. For lower values of $f$ ($<3.5$ mm), energy spread increases with time, attains a maximum value, and decreases again before it finally saturates at a constant value. However, the final saturated values of $\delta E_{fwhm}$ strongly depend on the laser focusing position. It is interesting to notice that the evolution of $Q_{fwhm}$, $\epsilon_n$, and $\delta E_{fwhm}$ follow a similar trend revealing a strong inter-correlation. The dependence of electron beam properties as a function of propagation distance is attributed to the combined effect of beam loading and evolution of the bubble size due to the self-focusing of the laser pulse. A detailed discussion has been provided in section \ref{effect_cn}.

The final beam properties of the trapped electrons after exiting from the plasma ($x>9.0$ mm) at time $\omega_{pe}t = 2212$ have been characterized for different simulation runs with changing values of $f$ and illustrated in Fig. \ref{beam_fc}. The energy histogram of the electron beam presented in Fig. \ref{beam_fc}(a) shows that for lower values of $f$, electron energy is spread over a larger range with lower values of maximum energy. Whereas, for higher values of $f$, energy distribution becomes narrower with higher values of maximum gained energy. The beam divergence representing the angular spread of the electron beam in the transverse plane also strongly depends on the laser focusing position, as can be observed from Fig. \ref{beam_fc}(b). It is seen that the angular spreading of the beam in the transverse plane becomes narrower with the increase of $f$. The peak value of gained energy of the electron beam and energy spreading as a function of $f$ has been depicted in Fig. \ref{beam_fc}(c). The peak energy ($E_{pk}$) increases monotonically with the increase of $f$, as also presented in Fig. \ref{beam_time_fc}. Whereas, the relative energy spread ($\delta E_{fwhm}/E_{pk}$) decreases with an increase of $f$ and finally it gets saturated beyond a certain value $f$ ($f\sim 3.7$ mm). The net absolute charge ($Q_b$) and normalized beam emittance ($\epsilon_n$) (calculated by considering all the trapped electrons) as a function of $f$ is shown in Fig. \ref{beam_fc}(d), which reveals that both $Q_b$ and $\epsilon_n$ decrease with an increase of $f$. It is to be noticed that the net charge of the beam can also be obtained from the energy histogram shown in Fig. \ref{beam_fc}(a). The total number of trapped particles can be obtained from the energy histogram by calculating the sum of the number of particles in each bin multiplied by the bin width (4 MeV, in our case). From this, one can calculate the charge (per unit length as for the 2D slab geometry) of the accelerated electron beam, which would be the same as presented in Fig. \ref{beam_fc}(d).

%~~~~~~~~~~~~~~~~~~~~~~~~~~~~~~~~~~~~~~~~~~~~~~~~~~~~~~~~~~~~~~~~~~~

\subsection{Effect of Nitrogen Concentration}\label{effect_cn}

\begin{figure}
  \centering
  \includegraphics[width=5.5in,height=4.0in]{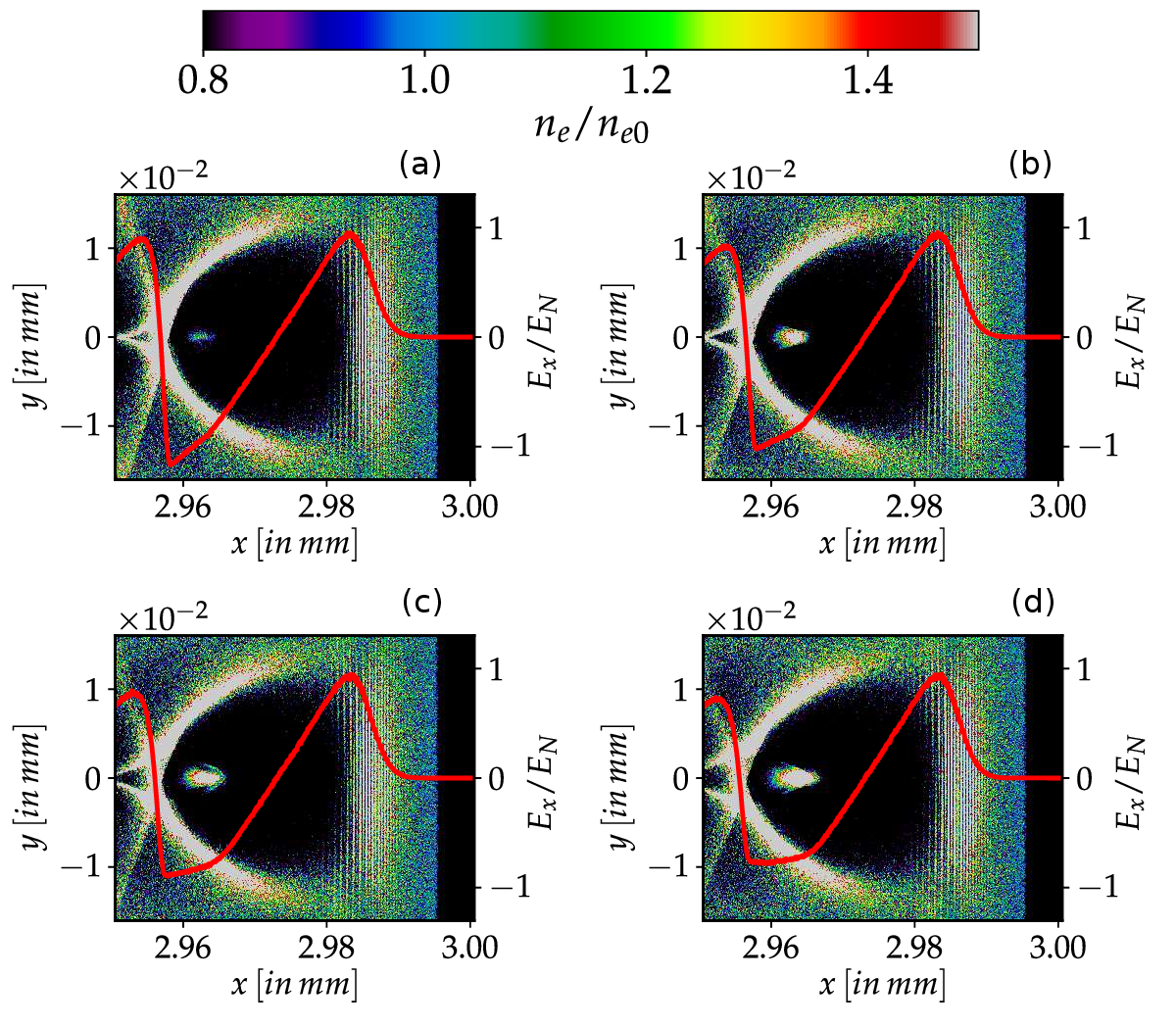}
  \caption{Electron density ($\rho$) distributions in the x-y plane at time $\omega_{pe}t = 715$ for simulation runs with N$_2$ concentration (a) $C_N = 7\%$, (b) $C_N = 10\%$, (c) $C_N = 12.5\%$, and (d) $C_N = 15\%$ are shown. Here, the red lines in each subplots represent the longitudinal electric field ($E_x$) for the corresponding cases. In these simulations, laser focusing position is kept fixed at $f = 3.5$ mm.}
\label{2d_dens_cn}
\end{figure}

\begin{figure}
  \centering
  \includegraphics[width=5.5in,height=4.0in]{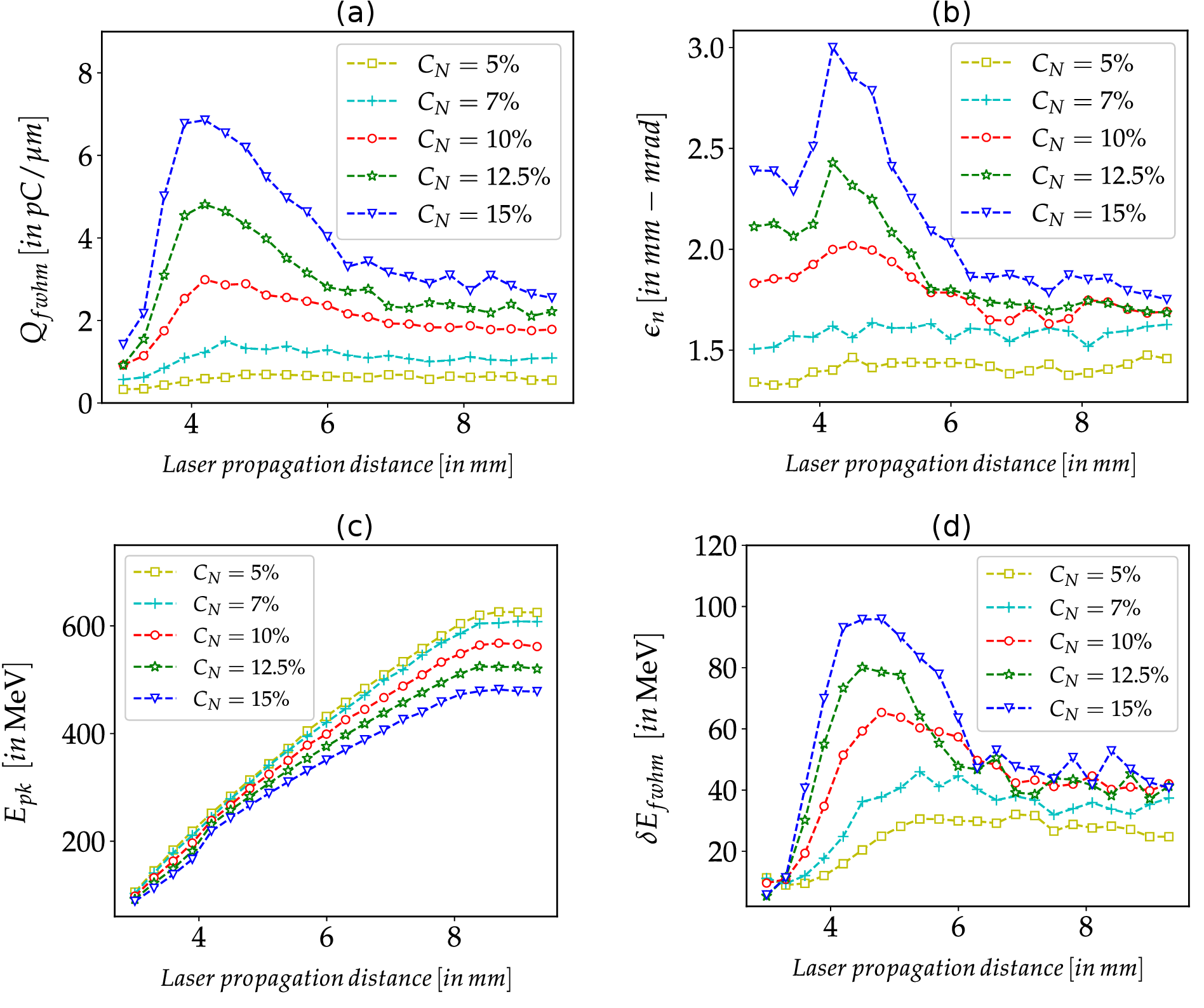}
  \caption{ Time history of the properties of electrons injected and trapped in the plasma wake structures behind the laser pulse is shown for different values of N$_2$ concentration ($C_N$). In subplots (a) and (b), the absolute values of charge ($Q_{fwhm}$) located within the FWHM of the energy distribution function of the trapped electron beam and corresponding normalized beam emittance ($\epsilon_n$) are shown as a function of time, respectively. Whereas, the time evolution of energy peak values ($E_{pk}$) and corresponding energy spread ($\delta E_{fwhm}$) are depicted in subplots (c) and (d), respectively.}
\label{beam_time_cn}
\end{figure}

\begin{figure}
  \centering
  \includegraphics[width=6.0in,height=3.0in]{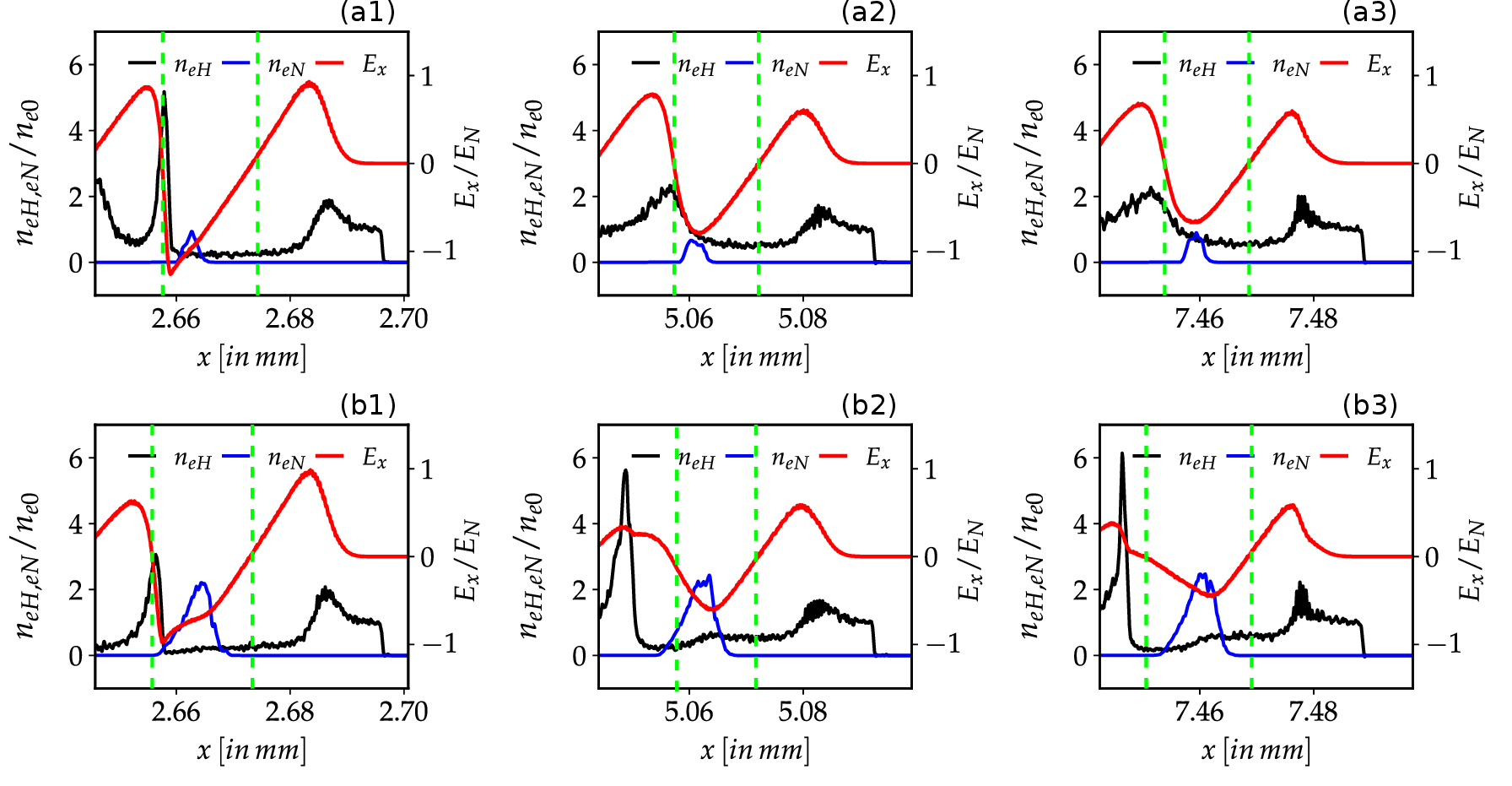}
  \caption{On-axis longitudinal electric field $E_x$ (red lines) and electron densities $n_{eH}$ (black lines), $n_{eN}$ (blue lines) originated from hydrogen, nitrogen neutrals, respectively, are shown at different times of evolution. Subplots (a1)-(a3) represent the case with nitrogen concentration $C_N = 5\%$. Whereas, for subplots (b1)-(b3), $C_N = 15\%$ was considered. Green dashed vertical lines define the locations where $E_x = 0$ inside the first bubble for each case.}
\label{bm_loding}
\end{figure}

\begin{figure}
  \centering
  \includegraphics[width=5.5in,height=4.0in]{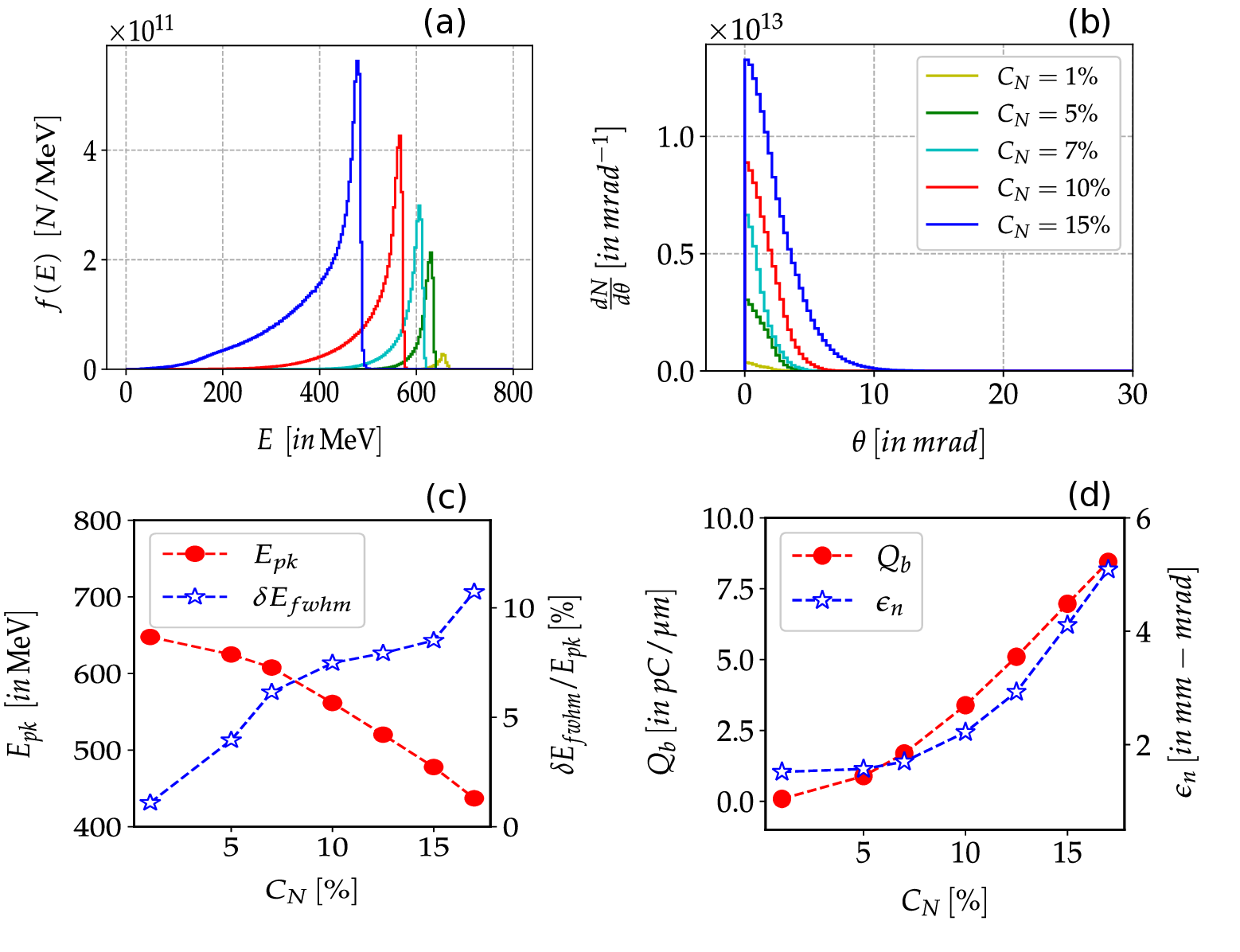}
  \caption{Energy histogram and beam divergence at a particular time $\omega_{pe}t = 2212$ for different simulations with changing values of N$_2$ concentration ($C_N$) are shown in subplots (a) and (b) respectively. Peak energy ($E_{pk}$) of electron beam, energy spread ($\delta E_{fwhm}$), beam charge ($Q_b$), and normalized beam emittance ($\epsilon_n$), with changing values of $C_N$ are depicted in subplots (c) and (d) at the same instant of time. Here, $Q_b$ and $\epsilon_n$ are calculated from the information of all the trapped electrons.}
\label{beam_cn}
\end{figure}

A set of simulation runs has also been carried out with changing values of N$_2$ concentration ($C_N$) to study its effect on the accelerated electron beam properties. In these simulations, the laser focusing position is fixed at $f = 3.5$ mm for a particular density configuration with $L = 1.0$ mm (as shown in Fig. \ref{dens_ini}). In these cases also, electrons ionized from N$_2$ are observed to be injected and trapped inside the wake structures as the laser propagates through the medium. The net charge density distributions in the x-y plane at a particular instant of time $\omega_{pe}t = 715$ have been shown in subplots (a)-(d) of Fig. \ref{2d_dens_cn} for different values of $C_N$. It is seen that an almost circular (spherical in 3D) bubble regime is created behind the laser pulse for all the cases. The bubble is a nonlinear stage of a wake structure that forms when a driver laser pulse is intense enough to expel almost all the electrons from near the laser pulse. The optimal condition \cite{lu2006nonlinear} to form a bubble is $k_pR \approx 2\sqrt{a_0}$, where $k_p = c/\omega_p$ is the plasma wave vector, $c$ is the speed of light in vacuum, $\omega_p$ represents the plasma frequency, $R$ defines the radius of the laser spot size, and $a_0$ is the normalized vector potential associated with the driver laser pulse. For our chosen values of simulation parameters, the value of $a_0$ should be approximately 3.1 to match this condition. Although, in our simulations, we have initially considered $a_0$ to be around 2.0 at the focus, due to the self-focusing effect inside the plasma, the value of $a_0$ reaches close to 3.0, satisfying the condition for the bubble regime to occur. The electrons are pushed forward and radially by almost the same strength of ponderomotive force as in our case $c\tau_{fwhm}\approx w_0$ incorporating the self-focusing effect. Consequently, the shape of the bubble becomes almost circular. The amount of charge trapped inside this wake structure increases with the increase of $C_N$. This is also apparent from the profile of the on-axis longitudinal electric field ($E_x$) shown by the red solid line in the subplots of Fig. \ref{2d_dens_cn}. A significant modification in the profile of $E_x$ is observed at the locations of trapped particles. This is a clear signature of the beam loading effect where an electric field is originated from the space charge of the trapped electrons and modifies the wake electric field profile. Fig. \ref{2d_dens_cn} demonstrates that the beam loading effect, hence the number of trapped electrons, increases with an increase of $C_N$. 

The temporal evolution of trapped electron dynamics and its characteristic properties are presented in Fig. \ref{beam_time_cn} for different simulation runs with changing values of $C_N$. We have observed that the in these cases also, the net charge ($Q_b$) of the trapped electrons remains constant throughout the evolution. However, the charge ($Q_{fwhm}$) within the FWHM of the energy distribution function of the trapped electrons is strongly dependent on the propagation distance, as can be seen from Fig. \ref{beam_time_cn}(a). For $C_N> 7\%$, $Q_{fwhm}$ first increases and attains a maximum value at a particular propagation distance. Later, after a certain distance of propagation, $Q_{fwhm}$ becomes constant. For $C_N \leq 7\%$, we have observed that $Q_{fwhm}$ increases first and then gets saturated after a certain distance of proportion. The normalized emittance ($\epsilon_n$) of the electron beam also strongly depends on the time of evolution for different values $C_N$, as seen from Fig. \ref{beam_time_cn}(b). For $C_N> 7\%$, the value of $\epsilon_n$ first increases with time, and then after a certain time, it decreases again before being saturated at constant values. However, for $C_N\leq 7\%$, beam emittance ($\epsilon_n$) almost remains constant throughout the evolution. The peak energy ($E_{pk}$) of the beam defining the energy acquired by the maximum number of trapped electrons is shown as a function of time in Fig. \ref{beam_time_cn}(c) for different values of $C_N$. It is seen that $E_{pk}$ increases monotonically (and almost linearly) with time for all the cases and finally gets saturated as the electron beams leave the plateau region of the second stage at $\omega_{pe}t\approx 2100$. However, it is observed that the final energy an electron beam gains is lower for the higher values of $C_N$. This is a consequence of the beam-loading effect. The amount of electrons injected and trapped inside the wakefield increases for higher nitrogen concentrations. Consequently, the accelerating field inside the bubble gets flattened due to the effect of the space charge electric field of the trapped electrons, as shown in Fig. \ref{2d_dens_cn}(d). As a result, the local electric field gradient, responsible for the acceleration of electrons, decreases for higher values of $C_N$, causing the peak value of final gained energy to be lower. In Fig. \ref{beam_time_cn}(d), the energy spread has been shown as a function of time for different simulation runs with changing values of $C_N$. The energy spread ($\delta E_{fwhm}$) is calculated from the full width at half maxima of the corresponding energy histogram of the trapped electron beam at different evolution times. It is interesting to observe that $\delta E_{fwhm}$ strongly depends on the time of evolution (i.e., propagation distance), and the value of $C_N$ significantly influences the nature of dependence. For example, with $C_N> 7\%$, $\delta E_{fwhm}$ initially increases with time, reaches a maximum value, and after a certain propagation distance, its value gradually decreases. Whereas, for the simulations with $C_N\leq 7\%$, the energy spread ($\delta E_{fwhm}$) of the trapped electron beams increases up to a certain time of evolution and finally gets saturated, as can be seen from of Fig. \ref{beam_time_cn}(d). In these simulations also, the evolution of $Q_{fwhm}$, $\epsilon_n$, and $\delta E_{fwhm}$ follow a similar trend.

As mentioned in section \ref{effect_fc}, the beam loading effect determines the nature of variation of electron beam properties as a function of propagation distance. The beam loading effect occurs when the space charge force of the accelerated electron bunch becomes significant enough to modify the structure of a wake and, consequently, the profile of the accelerating electric field. To demonstrate this phenomenon, we have evaluated 1D on-axis electron density ($n_{eH}$, $n_{eN}$) as well as the longitudinal electric field ($E_x$) at three different stages of evolution and depicted in Fig. \ref{bm_loding}. Here, we have considered two cases with $C_N = 5\%$ and $15\%$ and illustrated in Fig. \ref{bm_loding}(a1)-(a3) and Fig. \ref{bm_loding}(b1)-(b3), respectively. Fig. \ref{bm_loding}(a1)-(a3) shows that for $C_N = 5\%$, the charge of the trapped electron bunch (blue solid line) is not enough to make any significant modification of the wake structure. In this case, the size of the wake (i.e., in the first bubble) evolves slightly only due to the evolution of the laser spot size. Thus, for this case, the electron beam properties remain unchanged after a certain propagation distance, as shown in Fig. \ref{beam_time_cn}. However, for $C_N = 15\%$, the charge of the injected electron bunch becomes high and significantly modifies the bubble size as all as the profile of accelerating field $E_x$, as shown in Fig. \ref{bm_loding}(b1)-(b3). The positions of the density ($n_{eH}$) peak of the electrons originating from hydrogen (shown by black solid lines) indicate that the size of the bubble increases with the propagation distance. It is interesting to notice that there is an intermediate regime of propagation (Fig. \ref{bm_loding}(b2)) where a significant number of trapped electrons becomes out of phase with respect to the accelerating electric field (i.e., electrons lie in the regime where $E_x>0$). Thus, in this intermediate regime, energy spread $\delta E_{fwhm}$, and consequently $Q_{fwhm}$, $\epsilon_n$ increases, as shown in Fig. \ref{beam_time_cn}. However, as the laser propagates further, the wake structure and, consequently, the longitudinal electric field $E_x$ gets modified in such a way that all the trapped electrons get confined again within the accelerating phase of the field, as illustrated in Fig. \ref{bm_loding}(b3). Thus, electron beam properties, i.e., $\delta E_{fwhm}$, $Q_{fwhm}$, and $\epsilon_n$ decreases and become constant after a certain distance of propagation.

In these cases also, we have analyzed the properties of the electron beam at the final simulation time, i.e., after exiting from the plasma ($x>9.0$ mm) and shown in subplots (a)-(d) of Fig. \ref{beam_cn}. The energy histogram of the accelerated electron beam shown in Fig. \ref{beam_cn}(a) reveals that a quasi-monoenergetic beam is produced for all the values of $C_N$. It is also seen that the maximum energy gained by the electrons increases with the decreasing value of $C_N$. The energy distribution of trapped electrons also becomes narrower with the decrease of $C_N$. Fig. \ref{beam_cn}(b) shows that the divergence of the electron beam increases with an increase of $C_N$. However, beam divergence remains within $\sim 5$ mrad in all the cases. The peak energy ($E_{pk}$) of the beam decreases, whereas energy spreading ($\delta E_{fwhm}$) increases with an increase of $C_N$, as can be seen from Fig. \ref{beam_cn}(c). At the same time, it is observed that the net charge of the beam ($Q_b$) as well as normalized beam emittance ($\epsilon_n$) increases with the increase of $C_N$, as shown in Fig. \ref{beam_cn}(d). For the chosen values of parameters, we have observed that the $E_{pk}$ varies from 450-650 MeV with $\delta E_{fwhm}$ remains in between 1-10$\%$. The charge of the electron beam varies from 0.5-8.5 pC/$\mu$m in 2D slab geometry (approximately 5-85 pC in 3D if we assume the transverse width of the beam is 10 $\mu$m, which is
a typical beam width in LWFA). Whereas the values of beam emittance $\epsilon_n$ remain within the range 1.5-5 mm-mrad.

%~~~~~~~~~~~~~~~~~~~~~~~~~~~~~~~~~~~~~~~~~~~~~~~~~~~~~~~~~~~~~~~~~~~
\subsection{Effect of Density Profile}\label{effect_prfl}

\begin{figure}
  \centering
  \includegraphics[width=4.7in,height=2.5in]{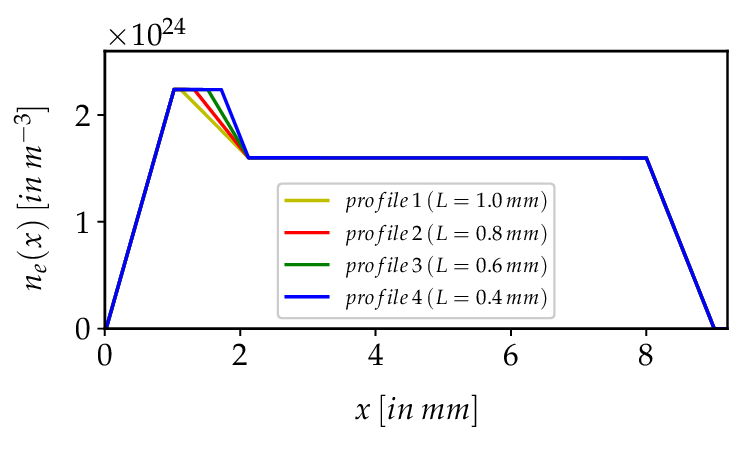}
  \caption{ The variations initial electron density along $x$ contributed from H$_2$ and N$_2$ (considering only five electrons from each nitrogen atoms) for four different simulation configurations with changing values of $L$ are shown. Here, the value of $L$ has been changed only by changing the position P2, keeping P1 and P3 fixed.}
\label{1d_ne_prfl}
\end{figure}

\begin{figure}
  \centering
  \includegraphics[width=5.5in,height=4.0in]{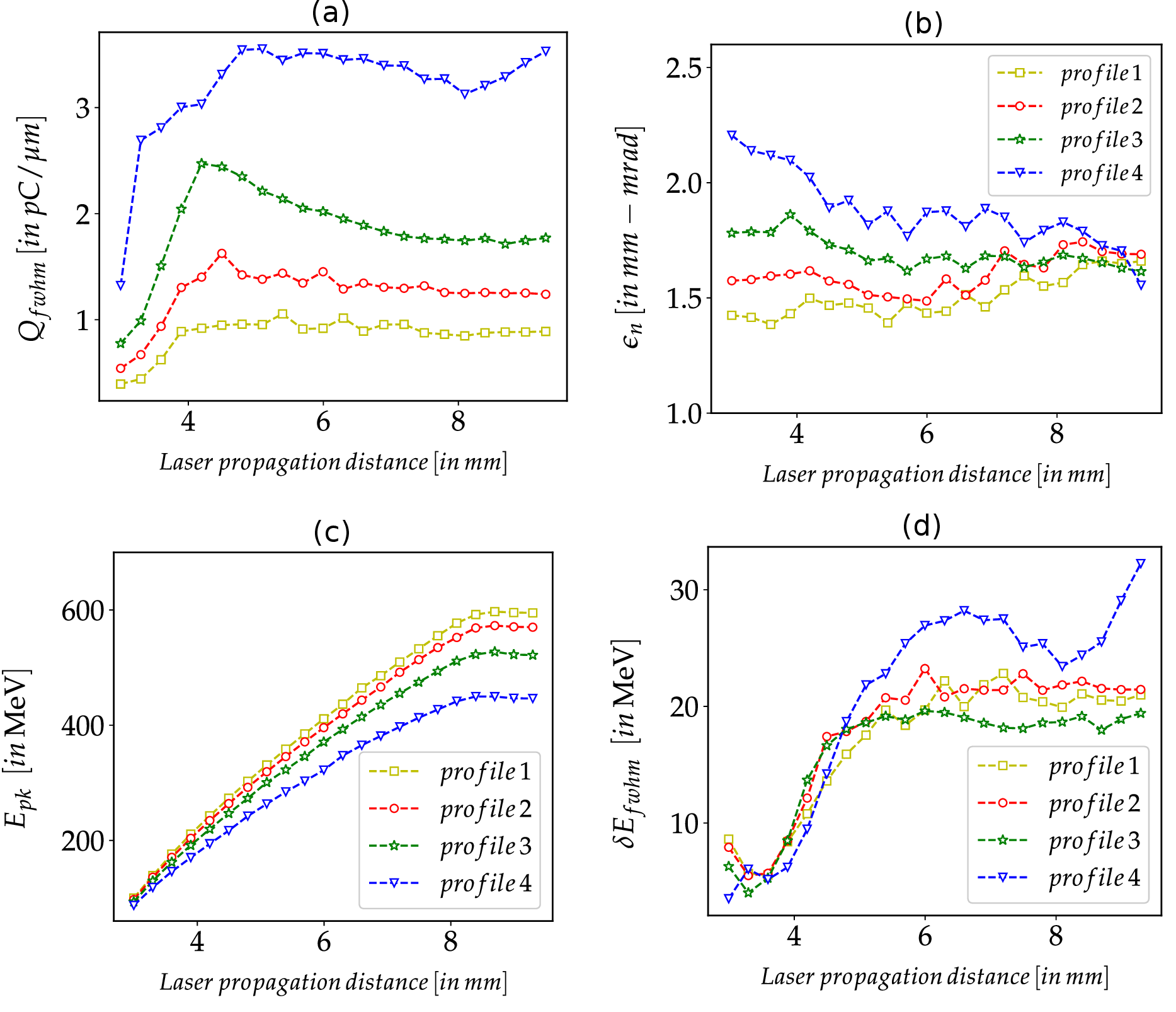}
  \caption{ The properties of electrons injected and trapped in the plasma wake structures behind the laser pulse as a function of laser propagation distance have been shown for different initial density profiles with changing values of $L$. In subplots (a) and (b), the FWHM charge $Q_{fwhm}$ and corresponding normalized emittance $\epsilon_n$ are shown as a function of propagation distance, respectively. Whereas, the evolution of $E_{pk}$ and $\delta E_{fwhm}$ are depicted in subplots (c) and (d), respectively.}
\label{beam_time_prfl}
\end{figure}

Our study also investigated the effect of the initial longitudinal density profile along the gas cell on the quality of accelerated electron beams. For this purpose, we have carried out simulations with changing values of $L$ with a fixed nitrogen concentration $C_N = 10\%$ and laser focusing position $f = 3.7$ mm. In particular, we have varied the position P2, keeping P1 and P3 at fixed positions, as illustrated in Fig. \ref{dens_ini}(b). The net electron (associated with the two and five electrons from each hydrogen and nitrogen atom, respectively) density ($n_e$) profile along $x$ for different values of $L$ has been illustrated in Fig. \ref{1d_ne_prfl}. As expected, the down-ramp gradient of $n_e(x)$ in the first stage ($x<2.12$ mm) increases with the decrease of $L$. In these cases also, we have analyzed several time-scanned properties of the trapped and accelerated electron beam and presented them in Fig. \ref{beam_time_prfl}(a)-(d). Fig. \ref{beam_time_prfl}(a) reveals that in all the cases, the absolute value of charge ($Q_{fwhm}$), located within the FWHM of the energy distribution function of the trapped electrons, increases up to a certain propagation distance and then gets saturate at particular values. We have observed a significant increase in electron beam charge with the decrease of $L$, as seen from Fig. \ref{beam_time_prfl}(a). These results are consistent with the density down-ramp injection mechanism, where it has been demonstrated that injection and trapping of electrons become more efficient with the increase of density gradient \cite{PhysRevE.58.R5257, PhysRevLett.86.1011, PhysRevLett.100.215004}. Moreover, in our case, the increase of beam charge is also attributed to the fact that the effective amount of N$_2$ becomes higher for lower values of $L$. The normalized beam emittance ($\epsilon_n$) calculated from the electrons within the FWHM of energy histogram almost remains constant throughout the propagation, as presented in Fig. \ref{beam_time_prfl}(b). It is also seen that $\epsilon_n$ has a higher value for the lower values of $L$. This is because $Q_{fwhm}$ is higher for lower values of $L$, and there is a direct correlation between $Q_{fwhm}$ and $\epsilon_n$. 

From Fig. \ref{beam_time_prfl}(c), it is seen that in these cases also, peak energy ($E_{pk}$) of the trapped electron beam increases monotonically (and almost linearly) with time, i.e., with the propagation distance. However, the final saturated value of $E_{pk}$ (i.e., after exiting from the density plateau in the second stage) is higher for higher values of $L$. This is also because of the fact that $Q_{fwhm}$ decreases with the increase of $L$. Thus, the modification of the accelerating field of the wake through the beam loading effect originating from the space charge of the trapped electrons becomes less significant for higher values of $L$. The energy spread ($\delta E_{fwhm}$) of the beam as a function of propagation distance has been shown in Fig. \ref{beam_time_prfl}(d) for changing values of $L$. It has been observed that in all the cases, $\delta E_{fwhm}$ first increases, and after a certain distance of propagation, its value gets saturated. This behavior also is consistent with the nature of the evolution of $Q_{fwhm}$ and can be easily understood from the point of view of the space charge effect of the trapped electrons. It is to be seen that in the case corresponding to profile4, $Q_{fwhm}$ increases again after 8 mm. This is consistent with the fact that for this case, $\delta E_{fwhm}$ increases again after an 8 mm propagation distance, as can be seen from Fig. 11(d). We have not observed any second injection at the exiting downramp, as the net charge of the beam remains constant throughout the evolution. However, the exact cause that triggers the increase of $\delta E_{fwhm}$ at the exit is still unclear. This may be attributed to the fact that for the case of profile4, the self-focusing of the laser pulse is higher as it propagates comparatively a longer distance in the higher density regime. Thus, even after propagating an 8 mm distance, the electric field inside the expanding bubble at the exiting downramp may affect the beam quality. In this set of simulation runs, for our chosen values of parameters, we have observed the generation of the electron beam with beam charge (2d) $Q_b$ varies in between 1-5 pC/$\mu$m (which is approximately 10-50 pC in 3D assuming a typical 10 $\mu$m transverse beam width) and normalized beam emittance ($\epsilon_n$) remains within 1.5-1.7 mm-mrad. The beam's peak energy $E_{pk}$ varies between 450-600 MeV with the corresponding relative energy spread 3-5$\%$.

%\begin{figure}
%  \centering
%  \includegraphics[width=5.7in,height=4.3in]{fig11.eps}
%  \caption{Energy histogram and beam divergence at a particular time $\omega_{pe}t = 2212$ for simulations with different initial density profiles (changing values of $L$) are shown in subplots (a) and (b), respectively. Peak energy ($E_{pk}$) of the beam, energy spread ($\delta E_{fwhm}$), net beam charge ($Q_b$), and normalized beam emittance ($\epsilon_n$), with changing values of $L$ are shown in subplots (c) and (d) at the same instant of time.}
%\label{beam_prfl}
%\end{figure}

%~~~~~~~~~~~~~~~~~~~~~~~~~~~~~~~~~~~~~~~~~~~~~~~~~~~~~~~~~~~~~~~~~~~
%~~~~~~~~~~~~~~~~~~~~~~~~~~~~~~~~~~~~~~~~~~~~~~~~~~~~~~~~~~~~~~~~~~~
\section{Summary and Conclusions}
\label{summary}

This study explores the parametric dependence of the properties of accelerated electron beams in a laser wakefield acceleration within a gas cell target. In particular, a systematic and comprehensive study to obtain high-quality electron beams in an LWFA based on a combination of ionization-induced injection and density down ramp injection mechanism has been carried out using Particle-In-Cell (PIC) simulations. The gas target considered in this study consists of two sections. In the first section, nitrogen-doped hydrogen gas was used. The first section is responsible for injecting and trapping the electrons in the wake structure. In the second section, pure hydrogen gas was considered. The second section provides the accelerating electric field to accelerate the trapped electrons further. Our study demonstrates the dependence of accelerated electron beam properties on laser focusing position, nitrogen concentration, and gas density profile. In particular, it has been explicitly shown that several important properties of an electron beam, e.g., beam charge, peak energy, relative energy spread, and beam emittance, are strongly influenced by the system parameters mentioned above. It has been observed that the energy peak value ($E_{pk}$) of the accelerated electron beam increases with the increase of laser focusing position ($f$). Whereas, beam charge ($Q_b$), normalized beam emittance ($\epsilon_n$), and relative energy spread ($\delta E_{fwhm}/E_{pk}$) decrease with an increase of $f$. Our simulation study also reveals that $E_{pk}$ decreases with an increase in nitrogen concentration ($C_N$). Whereas, beam charge ($Q_b$), normalized beam emittance ($\epsilon_n$), and relative energy spread ($\delta E_{fwhm}/E_{pk}$) increase with the increase of $C_N$. The optimal zones and suitable parametric regimes for beam delivery have been identified from the time analysis of the trapped electron beam properties. The effect of beam loading on the properties of accelerated electron beams at various stages of evolution has been analyzed and discussed. In our study, using the laser with 50 TW peak power and 30 $\mu$m (FWHM) spot size, an optimum parametric regime ($f = 3.3-3.7$ mm; $C_N = 5-12\%$) has been identified for which electron beams with peak energies of 500-600 MeV, relative energy spread less than $5\%$, normalized beam emittance around 1.5 mm-mrad, and 2D beam charge of 2-5 pC/$\mu$m (approximately 20-50 pC in 3D if we assume the transverse width of the beam is 10 $\mu$m) have been generated. These results suit nicely for the requirements of incoherent undulator physics and will be useful for the related experimental campaigns \cite{vishnyakov2023compact} at the ELI Beamlines facility in the future.

\section*{Data availability statement}
The data cannot be made publicly available upon publication because they are not available in a format that is sufficiently accessible or reusable by other researchers. The data that support the findings of this study are available upon reasonable request from the authors.

\section*{Acknowledgments}

This work was supported by the Ministry of Education, Youth and Sports of the Czech Republic through the e-INFRA CZ (ID:90254).
This work was also supported by the project “Advanced Research using High Intensity Laser produced Photons and Particles” (ADONIS) (CZ.02.1.01/0.0/0.0/16019/0000789) from European Regional Development Fund (ERDF). The authors gratefully acknowledge Prof. Sergei V. Bulanov for his valuable suggestions, which have significantly improved the quality of the manuscript. The authors thank Dr. Petr Valenta, a colleague from ELI Beamlines, for the helpful discussion and acknowledge the ELI Beamlines HPC facility for computational resources.
 
%\section*{Conflict of Interest}
% \noindent 
% Authors report no conflict of interest
 
%\section*{Data availability statement}
%\noindent
%The data can not be made publicly available upon publication because they are not available in a format that is sufficiently accessible or reusable by other researchers. The data that support the findings of this study are available upon reasonable request from the authors.

\section*{References}
\bibliographystyle{unsrt}
\bibliography{ref.bib}

%\begin{thebibliography}{999}

%\end{thebibliography}

\end{document}